# Laser self-injection locking to fiber Fabry-Perot resonator for frequency comb generation

Germain Bourcier, Stephane Balac, Julien Lumeau, Antonin Moreau, Thomas Bunel, Arnaud Mussot, Matteo Conforti, Vincent Crozatier, Olivier Llopis, and Arnaud Fernandez

*Abstract*—This study demonstrates that self-injection locking (SIL) of a distributed feedback (DFB) laser to a high-Q fiber Fabry Perot (FFP) resonator, fabricated with highly nonlinear fiber, allows optical frequency combs (OFC) generation with a laser power as low as 100 mW. More precisely, cavity soliton (CS) regime has been observed in this configuration, along with other types of combs. The laser stabilization using SIL is described. Then the system's behavior is analyzed through modeling the laser's dynamics and comparing the model results to experimental tuning curve measurements. Our findings highlight the critical role of the initial phase of the fiber link between laser and FFP in determining the stability and effectiveness of the locking process. We explore the dynamics of the nonlinear SIL process while varying the laser current, revealing the transition from modulation instability to chaotic comb states, and eventually to soliton formation as the system moves from an effective blue-detuned to an effective red-detuned regime. Notably, the inclusion of self-phase modulation (SPM) in the SIL model predicts accessibility of the narrow soliton existence range. These results highlight the potential of SIL in FFP resonators for low-power, stable OFC generation, offering a promising path forward for practical applications.

## I. INTRODUCTION

OPTICAL frequency combs were developed over two decades ago for optical metrology and optical clocks. These combs, with repetition rates ranging from MHz to THz, are now useful in a wide range of applications: from precision optical synthesizer [1], microwave high spectral purity signal generation [2] and telecommunications [3][4] to LIDAR [5][6] and spectroscopy [7]. Nonlinear Kerr cavities have emerged as a selected medium for the generation of OFCs and particularly combs based on cavity solitons. These solitons result from a double balance between anomalous group velocity dispersion (GVD) and Kerr nonlinearity on one side, and between losses and energy injection (with continuous-wave (CW) pumping) on the other side. The fiber Fabry-Perot is an alternative to microresonators and long-length fiber ring cavities. It allows for achieving high performance in terms of quality factor with resonators reaching up to $10^9$ [8], while remaining compact and easy to use thanks to Bragg mirrors deposited on the surface of FC/PC connectors using a physical vapor deposition technique to achieve a reflectivity beyond 99.8%. This type of resonator enables repetition rates of a few GHz and comb spanning up to 28 THz [9]. CS generation has already been demonstrated in FFP using CW [9] or pulsed pumping schemes. Stabilization is the key point for a successful generation of frequency combs. Stabilization management through the Brillouin effect [8][10] or using Pound-Drever-Hall (PDH) [11][12] has been successful to generate CSs in FFP, but all these demonstrations have required high-power CW pumping, up to 1 W. In this study we show that a self-injection locked diode laser with a high-Q FFP resonator made of highly nonlinear fiber allows, thanks to the low phase noise of the SIL laser [13], the generation of CS at much lower power threshold and using a low-cost DFB laser.

The paper is organized as follows. Section II presents the laser self-injection locking to a FFP using a model which is experimentally verified. In Section III, the model is extended to incorporate the nonlinearity of the resonator. Experimentally, we achieve the first demonstration of CS generation with less than 100 mW pump power in FFP resonators, verified through its hyperbolic secant spectrum and a beatnote at the repetition frequency with a narrow spectral linewidth leading to a phase noise below -80 dBc/Hz for an offset frequency above 10 Hz.

Manuscript received xxxxx xxxxxx; revised xxxx xxxxx; accepted xx xxxxxx. Date of publication xxxxx xxxxx; date of current version xxx xxxx. This work was supported in part by the Centre National d'Etudes Spatiales (CNES) and in part by the Agence de l'Innovation de Défense (AID). *(Corresponding author: Germain Bourcier).*

Germain Bourcier, Olivier Llopis, and Arnaud Fernandez are with the LAAS-CNRS, Université de Toulouse, CNRS, 31031 Toulouse, France (e-mail: germain.bourcier@laas.fr; llopis@laas.fr; afernand@laas.fr).

Germain Bourcier is with the CNES, F-31401 Toulouse, France (e-mail: germain.bourcier@laas.fr).

Stephane Balac is with the IRMAR, Université de Rennes, CNRS, 35042 Rennes, France (e-mail: stephane.balac@univ-rennes.fr).

Julien Lumeau, and Antonin Moreau are with the Aix Marseille University, CNRS, Centrale Med, Institut Fresnel, Marseille, France (e-mail: julien.lumeau@fresnel.fr; antonin.moreau@fresnel.fr).

Thomas Bunel, Arnaud Mussot and Matteo Conforti are with the University of Lille, CNRS, UMR 8523-PhLAM Physique des Lasers Atomes et Molecules, F-59000, Lille, France (e-mail: thomas.bunel@univ-lille.fr; arnaud.mussot@univ-lille.fr; matteo.conforti@univ-lille.fr).

Vincent Crozatier is with Thales Research and Technology, Palaiseau, France (e-mail: vincent.crozatier@thalesgroup.com).

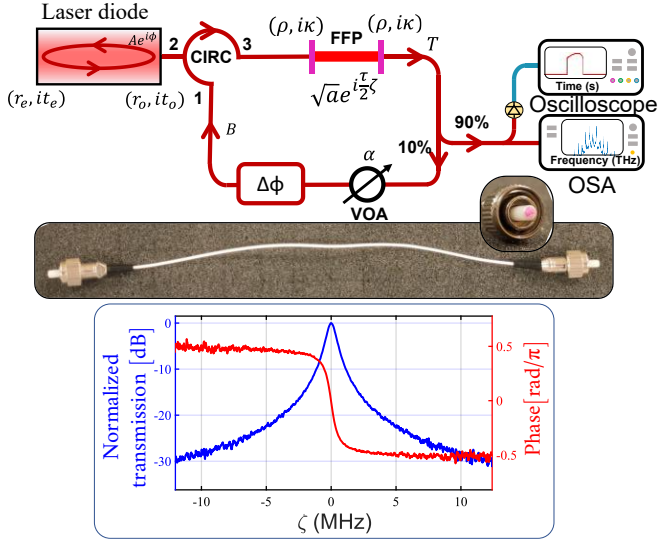

**Fig. 1.** Schematic representation of the SIL setup with a laser diode, a circulator (CIRC), a variable optical attenuator (VOA) and a phase shifter. Oscilloscope and optical spectrum analyzer (OSA) are used for measurements. The inset show a picture of the FFP resonator used in this experiment and its measured transmission function in amplitude and phase.

## II. SELF-INJECTION LOCKING

Optical feedback is an interesting approach for laser stabilization. Previously, laser SIL to high-Q microresonators has been used to obtain narrow-linewidth lasers, ultra-low-noise oscillators and soliton generation [14]. Wideband and high gain lock is possible by finding a compromise between strong feedback that minimizes noise and low feedback that decrease the possibility of system instability. The setup is presented in Fig.1. We use a variable optical attenuator (VOA) in order to adjust the feedback power. The FFP transfer function in transmission is used for the feedback, unlike in ring microresonators where the Rayleigh backscattering is used to stabilize the laser on the resonator thanks to its Lorentzian doublet transfer function in high-Q approximation. In a setup with a FFP, a long fiber loop including fiber circulator is then needed to feed the transmitted power back to the laser. The feedback of a filtered version of the laser allows the laser auto-refinement making the high stability. The phase induced by the fiber loop length, is a primary parameter influencing laser locking. The influence of this parameter had not been studied in a previous paper [13] where we demonstrated narrow-linewidth lasers using SIL to high-Q FFP. Therefore, a voltage-controlled phase shifter was added in the loop in order to control the phase and explore different stabilization regimes.

In this study, we take benefit of a FFP resonator we have developed to stabilize, through SIL, a distributed feedback (DFB) laser. The laser is a commercial device from Gooch & Housego, delivering up to 120 mW power near 1550 nm. The lack of isolator before the fiber coupling is mandatory. The FFP resonator is depicted in Fig.1. It is made from an optical highly nonlinear fiber (HNLF Thorlabs-HN1550P) of length $L = 20$ cm. Bragg mirrors are deposited to achieve a 99.86% power reflectivity over a 100 nm bandwidth. The nonlinear coefficient is $\gamma = 10.8$ W$^{-1}$km$^{-1}$, and the group velocity dispersion is in the anomalous regime and estimated to be $\beta_2 = -3.6$ ps$^2$km$^{-1}$. A finesse of 635 is obtained when characterizing the resonator through Spectro-RF method [15].

Geng *et al.* [16] have presented a detailed study of the impact of different parameters, such as injection ratio and initial phase, in the case of the stabilization of a fiber loop resonator. One of the interesting observations is the irregular evolution of the SIL due to random phase disturbance caused by external environment, such as temperature fluctuations. This is the reason why it is mandatory to put the whole setup in a thermal foam box. To validate this experimental setup and the proper functioning of the SIL, we first rely on measuring the noise of the laser stabilized on the FFP (Fig.2). A frequency noise floor far from the carrier below 1 Hz²/Hz is observed. In the vicinity of the carrier, between 10 Hz and 1 kHz, noise induced by vibrations limits the performance, especially since free spectral range (FSR) is very sensitive to vibrations in case of long fiber resonators [17].

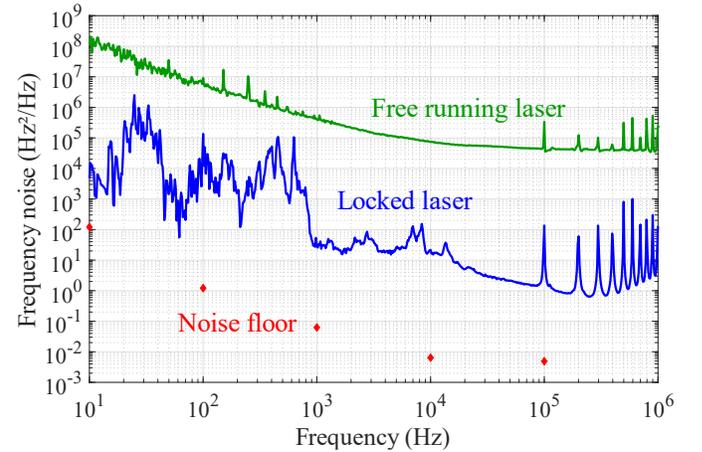

**Fig. 2.** Frequency noise of the free running laser (above, green) and the locked laser (blue). Red dots correspond to the measurement system noise floor [17].

We also carried out a laser current modulation experiment. By varying the laser current using a linear modulation ramp, we achieve a linear variation of the laser frequency from high to low frequencies. We define the Laser-FFP detuning parameter $\xi$ by the difference between the free running laser cavity (LC) frequency $\omega_{LC}$ and the closest FFP resonance $\omega_0$ such that $\xi = (\omega_0 - \omega_{LC})/2\pi$. The variation of $\xi$ with the laser current is almost linear. We also define the effective detuning $\zeta = (\omega_0 - \omega_{eff})/2\pi$ where $\omega_{eff}$ is the effective laser emission frequency which differs from $\omega_{LC}$ by the influence of the feedback on laser dynamics (Fig. 3). The "tuning curves" are defined as the dependence of $\zeta$ on $\xi$ [18][19]. When the laser frequency is far detuned from the resonance, the tuning curve follows $\zeta = \xi$. When close to resonance, the laser feedback induces laser SIL to the FFP so that $\omega_{eff} \approx \omega_0$ and remains robust to variations in $\xi$ within the locking range. We assume for simplicity a Fabry-Pérot cavity as a reduced model of the LC as depicted in Fig. 1. and we consider a single-frequency oscillation $\omega_{eff}$.

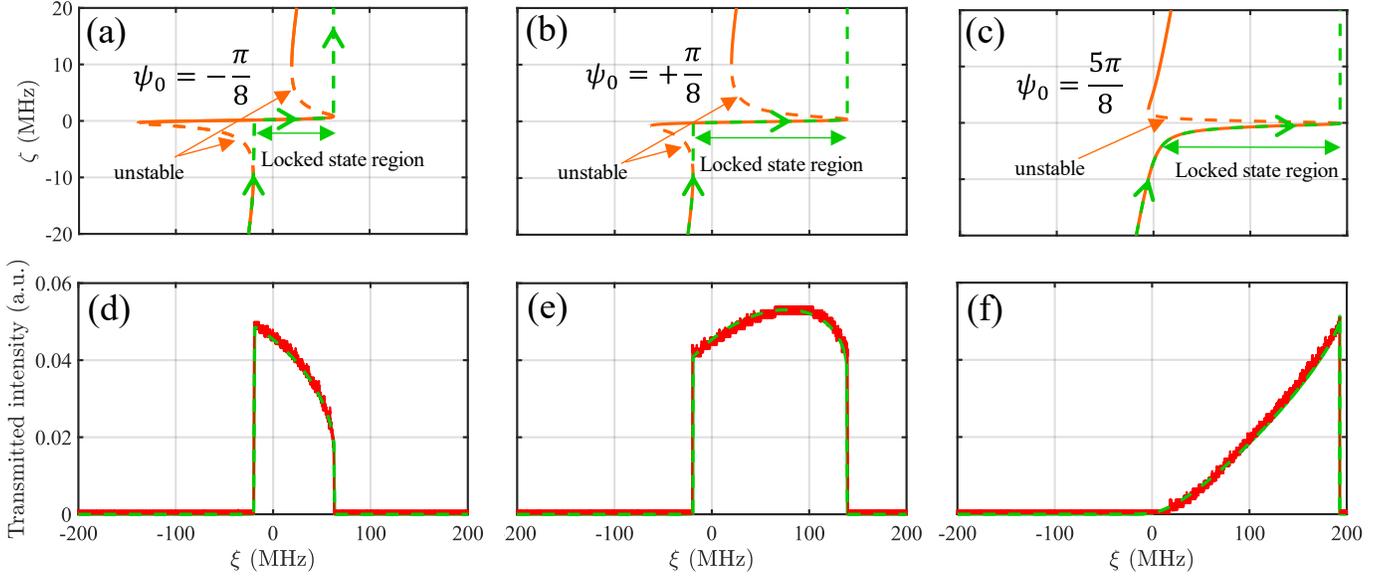

**Fig. 3.** (a, b, c) Simulated tuning curves (orange lines) and forward detuning path (green dashed lines) for different initial phase $\psi_0$. (a, d) $\psi_0 = -\pi/8$, (b, e) $\psi_0 = \pi/8$ and (c, f) $\psi_0 = 5\pi/8$. The dashed orange lines correspond to the unstable regions. (d, e, f) Simulated transmission of the FFP corresponding to the forward detuning path (green dashed lines) and experimental results for the corresponding initial phase (red lines).

This theoretical model stems from the work of Kondratiev *et al.* [18] that we derived for the FFP resonator with laser feedback in transmission. From this model we directly get a set of two equations, one for the amplitude $A$ and the second one for the phase $\phi$ of the LC field:

$$\dot{A} = \left(\frac{g(|A|^2)}{2} - \frac{\kappa_{LC}}{2}\right)A - \frac{\alpha t_0^2}{\tau_{LC} r_e r_o^2} A(t)|T(\omega_{eff})| \times \cos\left(\omega_{eff}\tau_s + \arg\left(T(\omega_{eff})\right)\right) \quad (1)$$

$$\dot{\phi} = (\omega_{eff} - \bar{\omega}_{LC}) + \alpha_g \frac{g(|A|^2)}{2} - \frac{\alpha t_0^2}{\tau_{LC} r_e r_o^2}|T(\omega_{eff})| \times \sin\left(\omega_{eff}\tau_s + \arg\left(T(\omega_{eff})\right)\right) \quad (2)$$

These equations have the form of Lang-Kobayashi (L-K) rate equations [21]. L-K equations describe light injection semiconductor laser in a single-mode model. In our case, the feedback depends on the frequency. The LC round-trip time is $\tau_{LC}$ and the amplitude reflection coefficients of the end and output laser mirrors are $r_e$ and $r_o$. $t_o = \sqrt{1-r_o^2}$ is the output transmission. $g$ is the material gain and $\alpha_g$ is the phase-amplitude coupling factor also called Henry factor [20]. $\tau_s$ is the round-trip time of the total fiber loop and $\alpha$ is the attenuation factor of the VOA. We define $\kappa_i$ as the photon decay rate due to intrinsic losses in the LC and $\kappa_c = 2(1 - r_e r_o)/(\tau_{LC} r_e r_o)$ as the decay rate due to coupling. $\kappa_{LC} = \kappa_i + \kappa_c$ is the total photon decay rate. $T(\omega_{eff})$ is the transfer function in transmission of the FFP resonator and $\bar{\omega}_{LC} = \omega_{LC} + \alpha_g \kappa_{LC}/2$ is the laser cavity frequency resonance. By multiplying the stationary form of Eq. (1) by $\alpha_g/A$ and subtracting it from the stationary form of Eq. (2), we get rid of $g(|A|^2)$, resulting in the following equation:

$$\xi = \zeta + \frac{\alpha K}{\tau_{LC}}|T(\zeta)|\sin\psi(\zeta) \quad (3)$$

where $K = t_0^2/(2\pi r_e r_o^2)\sqrt{1+\alpha_g^2}$ and $\psi(\zeta) = (\omega_0 - 2\pi\zeta)\tau_s + \arg(T(\zeta)) - arctan(\alpha_g)$ stands respectively as the combined coupling coefficient the phase delay.

We write the initial phase $\psi_0 = (\omega_0 - 2\pi\zeta)\tau_s - arctan(\alpha_g)$.

Unlike the other publications on injection locking in microresonators, we prefer to use the exact transfer function rather than its Lorentzian approximation as we don't need to simplify the equation to calculate Rayleigh backscattering. In the case of fiber Fabry-Pérot cavity, we use the transmission transfer function with $\rho$ and $\kappa$ the FFP mirrors reflectivity and transmissivity. $a$ represents the round-trip losses and $\tau$ is the round-trip time, $\tau = 2nL/c$, with $n$ the cavity fiber group index and $c$ the speed of light. The transfer function reads:

$$T(\zeta) = \frac{-\kappa^2\sqrt{a}e^{i2\pi\frac{\tau}{2}\zeta}}{1 - a\rho^2 e^{i2\pi\tau\zeta}} \quad (4)$$

For our $L = 20$ cm resonator, the parameters extracted from transmission measurement, see Fig. 1 inset, lead to a reflectivity $\rho = 99.93\%$, a transmissivity $\kappa = 3.74\%$, and round-trip losses $a = 0.995$. We explore analytically three initial phase values and carry out a forward detuning corresponding to an increase in the laser current and therefore to a detuning from high to low LC frequencies. Tuning curves are calculated in this case, as drawn in orange in Fig. 3. The middle branch is unstable in the three-solution region [18] (dashed orange line). We start the scan far from the resonance where the effective frequency follows linearly the LC frequency. Approaching the FFP resonance, it goes along the stable branch (solid orange line) until a turning where the frequency directly jumps to the inner stable branch (cases $\psi_0 = \pm\pi/8$) taking the path plotted in green dashed lines. Fig. 3(d-f) also shows the evolution of the intensity transmitted by the resonator during this scan with the dashed green curves. Jumps can clearly be seen as the transmitted intensity goes directly from zero to high value. Increasing $\xi$ further can lead to a laser unlock corresponding to a jump in frequency far from the resonance and seen as a sudden drop in transmitted intensity. The locking range correspond to the Laser-FFP detuning range $\xi$ associated to an effective detuning $\zeta$ on a curve close to a plateau like the inner stable branch. The lock state region is the region in the locking range corresponding to a one-way scan (described by double arrow in Fig. 3). This region is different for a forward and a backward scan. Fig. 3 only shows forward scan. The third initial phase case, $\psi_0 = 5\pi/8$, has no inner stable branch in the locking curve and do not give access to the tip of the resonance but the laser is locked to the side of the resonance. This case is noticeable since there is no frequency jump towards resonance clearly seen in the two others transmitted intensity curves. However, it is more interesting for many applications to be locked at the top of the resonance and it is well known that the case $\psi_0 = 0$ corresponds to the flattest inner curve and therefore to the greatest stability of the effective frequency [18]. Fig. 3(d-f) show scans performed experimentally by varying the phase of the fiber loop using the phase shifter which is equivalent to modifying the value of $\tau_s$. We measure the power transmitted by the FFP during laser frequency detuning. The initial phase situations $\psi_0 = \pm\pi/8$ ; $5\pi/8$ were obtained, thus making it possible to validate the fidelity of the analytical model to the reality of the experiment when the intensity is low enough not to generate nonlinear effects. The "tuning curve" undergoes a change with initial phase $\psi_0$ associated with the length of fiber in the loop and interference in the laser [16][18].

### III. SIL TO A NONLINEAR FFP

Numerous studies on integrated ring resonators utilize the SIL process to generate frequency combs, and particularly cavity solitons. This process is often referred to as "soliton SIL". It is widely used for its ability to stabilize the pump laser without the need for a complex electronic stabilization system, such as PDH [11][12]. The locking bandwidth of the system can also be significantly larger, reaching up to the FSR in the case of a strong feedback from the Laser-FFP detuning point of view, while it is typically limited to a few MHz for a PDH stabilization. In terms of noise reduction bandwidth, PDH rejects noise up to around 10 kHz [15][17], whereas the SIL technique extends the noise reduction bandwidth beyond the MHz range [13]. Consequently, the stability performance provided by SIL is exceptional compared to PDH. However, the disadvantage of SIL is its sensitivity to environmental disturbances, especially due to the long fiber lengths involved, unless the system is perfectly thermally and mechanically isolated. This is not the case with PDH, which is inherently more robust against such environmental factors. Additionally, SIL allows the use of lasers whose characteristics would normally be insufficient for comb generation and are low cost compared to commercial highly stable and tunable lasers. The stabilization process in the absence of the Kerr effect has been described in part II.

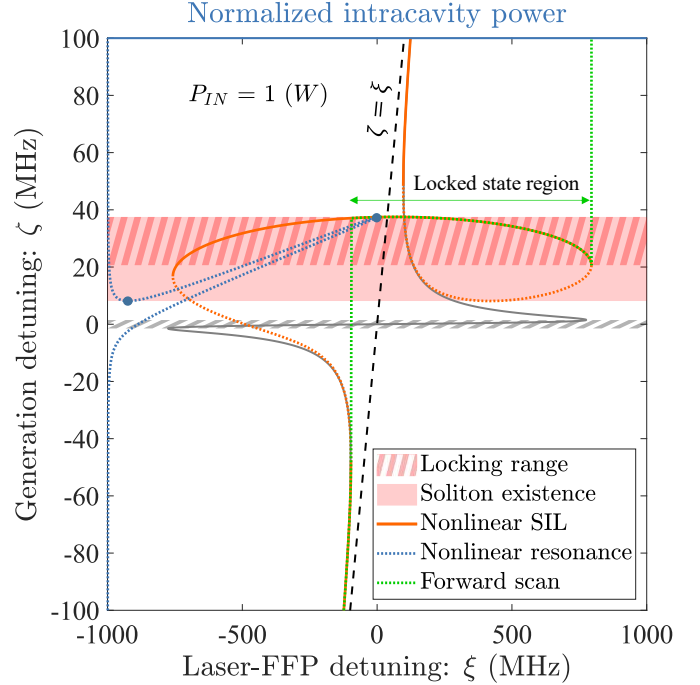

**Fig. 4.** Simulated tuning curves including self-phase modulation with an input power of 1 W (orange lines) and forward detuning path (green dashed lines) for initial phase $\psi_0 = 0$. The grey line represents the simulated tuning curves without self-phase modulation. The corresponding tilted resonance shown in blue pointed line (Normalized intracavity power as a function of $\zeta$).

The presence of the Kerr effect, particularly self-phase modulation, results in a resonance shift proportional to the intra-cavity power leading to the tilted resonance depicted in blue in Fig. 4. Locking directly within the soliton existence region has been experimentally demonstrated in the context of ring resonators [14][22]. This not clearly understood process has been analytically explained by incorporating self-phase modulation into the SIL model [19]. The usual process involves detuning from blue to red (forward scan), allowing access to the soliton step. The inner branch of the tilted resonance is typically inaccessible due to bistability and hysteresis behavior during forward and backward scans [23]. In contrast, SIL does not provide qualitative access to the soliton step, as falling out of resonance causes the laser to unlock, to jump in frequency far from the resonance and to lose its high quality (low phase noise) provided by the SIL. However, the rapid adjustment of the effective laser emission frequency $\omega_{eff}$ enables the laser to achieve a locked state region within the soliton existence range (pink area in Fig. 4).

The soliton existence range, in the case of ring resonator, is defined by a lower bound of detuning corresponding to the onset of the multi-stability region and an upper bound directly proportional to the intra-cavity power and slightly higher than the detuning corresponding to the resonance peak [24]. In the case of laser SIL to resonators, the locking range restricts the soliton existence to the multi-stability region [25][26]. The soliton existence detuning range is depicted by the light red area in Fig. 4 whereas the locking range corresponds to the dashed red area. It depends on the initial phase which is chosen as $\psi_0 = 0$ in Fig. 4. Incorporating SPM involves shifting the detuning according to the following relations:

$$\bar{\xi} = \xi - 6L\gamma P_{IC}/2\pi\tau \quad (5)$$
$$\bar{\zeta} = \zeta - 6L\gamma P_{IC}/2\pi\tau \quad (6)$$

where $P_{IC}$ is the intra-cavity power. The nonlinear tuning curves become.

$$\bar{\xi} = \bar{\zeta} + \frac{\alpha K}{\tau_{LC}}|T(\bar{\zeta})|\sin\psi(\bar{\zeta}) \quad (7)$$

where $T(\bar{\zeta})$ can be given by relation (4) or calculated using the stationary Lugiato-Lefever equation (LLE) [27].

Relation (7) describes tuning curve associated with the nonlinear SIL and plotted in Fig. 4 with orange curve. The forward laser scan (blue to red) results in a specific locked state region in the locking range (green dashed lines). Grey lines and dashed area show respectively the tuning curves and the locking range with the linear SIL. It is worth noting that a backward scan also allows access to a locked state region within the soliton existence range, which is not possible with the conventional method.

The results presented here correspond to real-world parameters for the FFP resonator employed in our experiments, with the exception of the power injected into the resonator $P_{IN}$ for Fig. 4. The power of 1 W was chosen to produce a clear and easily interpretable figure due to the strong nonlinearity applied. However, this power is not attainable in practice with our DFB laser, which is limited to 120 mW. The initial phase $\psi_0$ can be adjusted by adding a phase shifter before the injection feedback of the signal filtered by the resonator into the laser. Changing the initial phase directly impacts the locking state region. Fig. 5 shows the locking curves for different initial phase values. The 2D plot is overlaid on the soliton existence range (pink band in Fig. 5), showing that only specific initial phase values allow access to it. It is crucial to emphasize that an initial phase between $-\pi/3$ and 0 is particularly well-suited for the generation of CS. This phase range plays a key role in achieving the desired soliton state, highlighting the importance of a stable phase during soliton formation. It is important to note that the single-mode fiber used are not polarization-maintaining. Polarization is optimized with respect to injection into the resonator to maximize the transmission and consequently, the intra-cavity power as well. To optimize feedback and therefore the locking range, it is also important to maximize the re-injection on laser operating polarization.

The experiment previously described for the linear resonator, i.e., at low power, is repeated with a power level exceeding the threshold for parametric oscillations [27]. The power is set between 80 and 120 mW. The laser current is manually increased to perform a forward detuning scan. At a certain detuning point of -20 MHz (see Fig. 6(b)), the laser diode becomes self-injection locked to the resonator. Once it is locked, the laser linewidth reduces from about a few hundred kilohertz to a few hundred hertz, or even less than one hertz in terms of instantaneous linewidth [13], which increases the efficiency of the optical source to generate low noise combs [28]. The lock is initially achieved in a blue-detuned region relative to the effective resonance. It is thus possible to remain below the modulation instability (MI) threshold [27] depending on the laser-FFP detuning and on the initial phase. But increasing the detuning or decreasing the initial phase will progressively increase the power and lead to generation of MI. For example, Fig. 6(b) shows a lock close to the top of the effective resonance that led directly to primary comb (Fig. 6(c)) as the input power is only 100 mW. As the injection current continues to increase it is followed by various stages of MI chaotic combs gradually filling all the cavity modes (Fig. 6(d)).

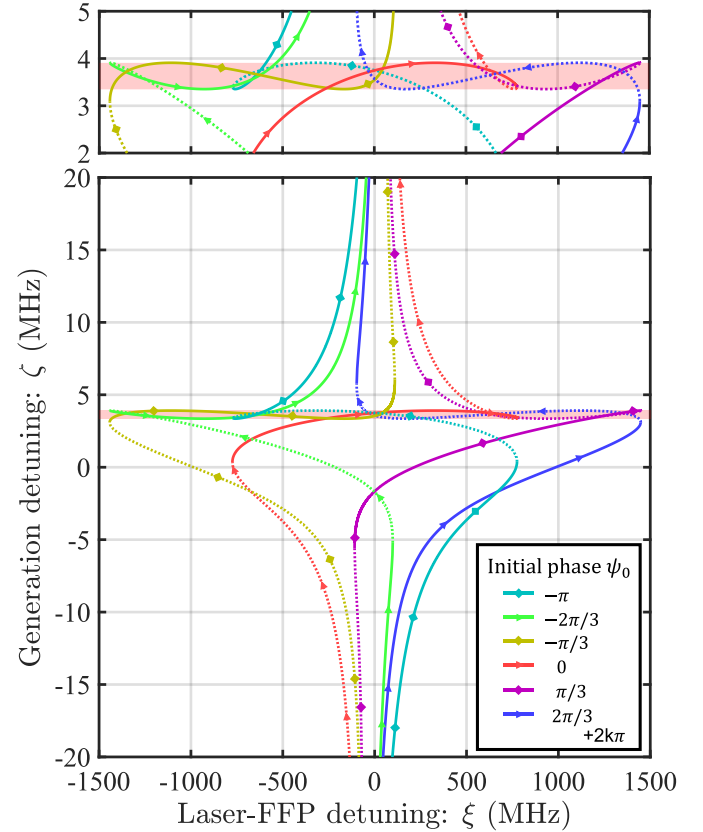

**Fig. 5.** Locking curves for different initial phases $\psi_0$ and $P_{IN} = 0.1\ W$. Plot of the locking curves overlaid on the soliton existence range (pink band). The inset shows a zoom on the soliton existence range.

After passing a maximum transmitted power the chaos gradually diminishes (Fig. 6(e)), ultimately leading to the formation of a CS state (Fig. 6(f)). Fig. 6(b) is the transmitted power when doing the detuning scan from blue to red frequencies. We observe a progressive decrease in transmitted power, indicating that the average power diminishes as we transition from chaotic combs (Fig. 6(d)(e)) to a soliton crystal (orange curve in the inset of Fig. 6(h)), then to multiple solitons (Fig. 6(g)(h))), and finally to a single soliton (Fig. 6(f)). With a pump power of 100 mW, the soliton existence range is very narrow. The soliton's lifetime is highly dependent on phase and temperature fluctuations, making it unstable over time. Temperature fluctuations are significantly mitigated by the foam box. The only remaining effects are the variations in power within the FFP and the induced thermo-optical effect. However, this does not present a major issue as the laser power is only 100 mW, thus resulting in minimal variations. Moreover, the SIL mechanism is highly robust against these fluctuations. The short fiber lengths across all components also significantly improve system stability, allowing us to measure the beatnote of combs (Fig. 6(g)) and their phase noise (Fig. 6(h)). This enabled us to observe a variety of combs, ranging from multiple solitons to single soliton, as well as soliton crystals. An undeniable advantage is the ability to tune the laser along the inner branch of the locking curve, making it possible to access a solitonic state and then retrace steps to reach multiple soliton states, chaotic combs, and even return to the primary comb state (Fig. 6(c)), simply by manually reducing the laser current.

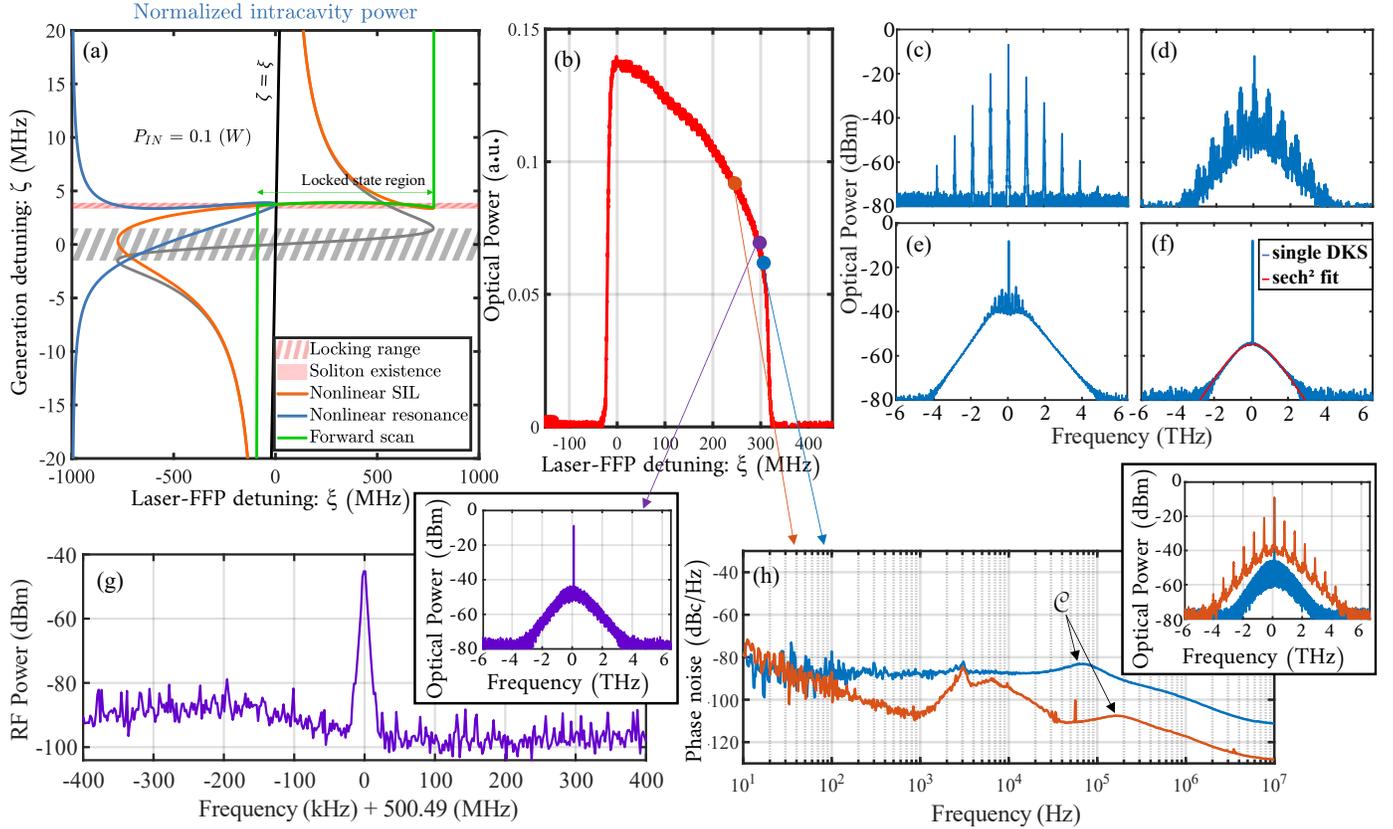

Fig. 6. (a) Simulated tuning curves with an initial phase $\psi_0 = 0$ matching with the experimental transmission in (b). Purple point corresponds to the DKS state presented in (g). (c, d, e, f) Experimental optical spectrum of comb states for different detuning frequencies with an input power of 100 mW. (g) RF beatnote of this multiple dispersive Kerr soliton (DKS) at the repetition frequency. The inset show the optical spectrum of the comb. (h) RF phase noise measurement of multiple solitons (blue curve) and soliton crystal (orange curve), and their respective optical spectra (inset).

It was not possible to measure the phase noise of a single soliton due to the low power in the comb lines. However, the phase noise of a soliton crystal and a multiple soliton state was measurable (Fig. 6(h)). The phase noise is lower in the case of the soliton crystal, dropping below 120 dBc/Hz at higher offset frequencies. Analysis of these phase noise curves reveals two bumps. The first, between $10^3$ and $10^4$ Hz, remains unidentified and will require further investigations, while the second, at frequencies above several tens of kHz, shown with arrow on Fig. 6(h), is characteristic of the resonant response $\mathcal{C}$ of the resonator in a red-detuned regime [23][29]. The multiple soliton state corresponds to a lower average power compared to the soliton crystal, and the bump appears further left on the phase noise curve. This seems to indicate that the generated laser frequency is closer to the cold cavity resonance but this will also require further investigation.

To our knowledge, this result represents the first demonstration of achieving a CS in a centimeter-scale FFP resonator with a pump power as low as 100 mW. Moreover, the use of a simple DFB laser, without any EDFA to boost the power, is a low cost and efficient approach to get optical micro-combs in these resonators.

IV. CONCLUSION

In conclusion, this study demonstrates the successful generation of cavity solitons within a fiber Fabry-Perot resonator using a self-injection locked distributed feedback laser at power levels as low as 100 mW. The findings highlight the potential of FFP resonators as a compact and efficient alternative to microresonators for low-power OFC generation. By integrating the nonlinearity of the resonator into the SIL model with the pump self-phase modulation, we have shown the possibility of accessing the soliton existence range. We experimentally achieved soliton formation, a significant step forward in practical applications of frequency comb technology. This work paves the way for further advancements in compact, low-power OFC systems, with implications across various fields.